# Compilation of mathematical expressions in Kotlin

Iaroslav Postovalov
*JetBrains Research (Nuclear Physics Methods group)*
Novosibirsk, Russia
ORCID: 0000-0002-7221-365X
postovalovya@gmail.com

*Abstract*—Interpreting mathematical expressions at runtime is a standard task in scientific software engineering. There are different approaches to this problem from creating an embedded domain-specific language (eDSL) with its own parser and interpreter specifically for that task, a full-fledged embedded compiler. This article is dedicated to a middle-ground solution implemented in the KMath library, which uses the Kotlin object builder DSL and its own algebraic abstractions to generate an AST for mathematical operations. This AST is then compiled just-in-time to generate JVM bytecode. A similar approach is tested on other Kotlin platforms, where its performance is compared across a variety of supported platforms, and we show JVM and JavaScript.

## I. Introduction

A common task in scientific software development is the dynamic interpretation of mathematical expressions, where an expression, either defined as source code or an external string, must be evaluated at runtime.

There are broad two approaches to dynamic execution: pure interpretation is more portable but typically slower, and dynamic compilation (i.e. JIT [1], or just-in-time compilation) is often faster but more complex. As the goal of this research is to provide a universal and high-performance dynamic execution engine, pure interpretation does not satisfy our requirements, and JIT compilation is too complex, requiring many tradeoffs between the level of optimization and compiler performance.

Various runtimes offer competitive performance and provide a mechanism to load the intermediate representation (IR) dynamically, among them, the well-known *Java Virtual Machine* (JVM). In addition to the Java language itself, the JVM supports a variety of other languages, including a modern, expressive language called Kotlin [2].

A reasonable question to ask is: why implement a custom compiler infrastructure instead of using an existing solution? For example, Kotlin provides a scripting engine for compiling expressions on-the-fly. However such general-purpose compilers are limited by their size and complexity (e.g., `kotlin-compiler-embeddable` is around 40 MiB), and the compilation performance is usually slow.

The alternative described in this paper is implemented in a library called *KMath* [3], [4], a Kotlin-based mathematical library, which implements a generic set of elements and abstract algebraic operations over them.

The approach taken by KMath offers a number of benefits for defining and evaluating mathematical expressions. In particular, users can easily implement expressions on double-precision floating-point numbers and define custom operators on all supported algebraic structures.

## II. KMath Principles

Mathematical operations in KMath are separated from mathematical objects. To perform an operation, e.g. `+`, one needs two generic objects of type `T` and a polymorphic algebraic context parameterized by `T`, e.g. `Space<T>`. Separating operations from objects has several advantages:

- Multiple operations may be valid depending on the application. For geometric applications, a matrix has an elementwise sum but no corresponding operator for multiplication. While NumPy semantics allow elementwise addition, multiplication, and other operators, mixing these operations in a single type can lead to confusion and complicate implementation.
- Multiple implementations of the same operation may be possible. For example, *KMath* supports binding to external libraries that may be used interchangeably.
- The context (called `Algebra` or algebraic context) could store information required to provide additional runtime guarantees. For example, it is possible to guarantee only a specific shape of n-dimensional arrays is valid in a given context or lexical scope. [5]

Mathematical contexts have the following hierarchy:

$$\texttt{Field <: Ring <: Space <: Algebra}$$

These interfaces loosely fulfill the standard mathematical definition designated by their name:

1) `Space` defines +, additive identity (i.e., 0), and ×.
2) `Ring` adds a multiplicative identity (i.e., 1).
3) `Field` defines a multiplicative inverse, i.e. ÷

A typical implementation of `Field<T>` is the `RealField` which works on doubles, and `VectorSpace` for `Space<T>`. In some cases, the algebra context can hold additional operations such as `exp` or `sin`, which are inherited from the appropriate interface.

## III. DESIGN OF THE EXPRESSIONS API

The KMath abstract algebra offers additional benefits for expression compilation. One can create a generic expression that uses only operations, provided by an `Algebra` contract, and then use a specific `Algebra` instance to perform operations and compute the value of the expression. Still, it requires some additional work to enable dynamic expression compilation.

The first major change of the KMath core API was the addition of dynamic operation dispatching to the primary marker interface `Algebra<T>`. As no algebraic context in KMath requires ternary operations, only `unaryOperation` and `binaryOperation` methods were added to call an operation dynamically by name and operands. The `symbol` method was introduced so that `Algebra<T>` could declare constants like the imaginary $i$ in the complex number system. Also, `unaryOperationFunction` and `binaryOperationFunction` companion functions were added, which both return an object of the Kotlin function type instead of the value of the operation.

The second stage of implementation involved submitting expressions as an entity in KMath: in the current API, instances of `Expression` should declare a function `invoke` which takes bindings and evaluates the expression. The most basic implementations of `Expression` are the so-called *functional expressions*, which are organized as a tree of `Expression` objects, which we describe below.

## IV. THE MST STRUCTURE

The *MST* (Mathematical Syntax Tree) is a primitive abstract syntax tree representing a language of mathematical expressions. Loosely, it consists of the following grammar:

$$\begin{aligned}
\langle\text{terminal}\rangle &\models \langle\text{symbol}\rangle \mid \langle\text{number}\rangle \\
\langle\text{unaryEx}\rangle &\models \langle\text{unaryOp}\rangle\langle\text{mst}\rangle \\
\langle\text{binaryEx}\rangle &\models \langle\text{mst}\rangle\langle\text{binaryOp}\rangle\langle\text{mst}\rangle \\
\langle\text{mst}\rangle &\models \langle\text{terminal}\rangle \mid \langle\text{unaryEx}\rangle \mid \langle\text{binaryEx}\rangle
\end{aligned}$$

KMath provides three ways to obtain MST instances:
1) Parse a string using a more specific grammar (stored in `ArithmeticsEvaluator.g4` file in [4]) that can produce a certain set of MST nodes from strings like `sin(x) ^ 2 + 25`.
2) Construct it explicitly.
3) Use a special KMath context where all the operations create an MST.

The MST can be interpreted recursively, although this is the slowest method of evaluation, as will be shown in Section IX. KMath provides such an interpreter for three reasons: dynamic compilation is restricted in several VM environments, is not implemented in Kotlin/Native, and the interpreter is useful for testing and debugging.

MST APIs are connected to the `Expression` API via the `MstExpression` class, which consists of a pair of an MST node and a reference to an algebraic structure. The only difference between `MstExpression` and direct MST interpretation is that in the `Expression` implementation, symbolic nodes are loaded not only with the constants and literals of the corresponding algebraic context but also with the expression symbols.

Four other approaches for `MST` translation were considered, two of which were performant and could compile `MST` instances for any algebraic structure — both user-declared and those loaded from within any KMath module.

## V. JAVA CLASS GENERATION

The goal of `MST` compilation to Java bytecode is to fetch and load Java classes dynamically from an `MST` instance. The generated class should implement the `Expression` interface with valid type parameters, be consistent with the interpreter and delegate all operations directly to a KMath algebraic structure to be universal.

*ObjectWeb ASM* [6] was chosen as a bytecode manipulation framework, as it was considered to be the most lightweight and widely-used in the compiler industry.

This approach presented two major problems.

The first is boxing. Boxing and unboxing type conversions [7] are often performed on the JVM due to type erasure but degrade calculation performance. Since `Expression` is a generic interface, boxing conversions are performed there so the number of these conversions can be minimized. It is possible to optimize out boxing with escape analysis and scalar replacement, but the performance of generated expressions on alternative JVM implementations such as *GraalVM* [8] must be investigated further.

The second problem concerns how to acquire a method signature that will call the needed algebraic operation. Four potential solutions were considered:
1) The method to invoke is searched within the given `Algebra` object with Java reflection.
2) Rather than calling algebraic operation functions directly, the `unaryOperation` and `binaryOperation` routines are called each time.
3) Direct calls are inserted, but only if the user provides a dictionary mapping `Algebra` operation identifiers to method signatures.
4) `unaryOperationFunction`/`binaryOperationFunction` are used, and the functions they return are stored within the expression object.

Each of the four options has advantages and disadvantages which are contrasted in Table I.

Two attempts were made to implement Java bytecode generation. The first used reflection-based lookup; however, this approach failed in cases when there was a mismatch between the operation name and the name of the function retrieved, or when the semantics of the operation did not match those of the corresponding function. Implementing this algorithm proved cumbersome, requiring coercion of stack values and careful use of reflection.

In the second attempt, Option 3 was accomplished— function type objects were collected from `unaryOperationFunction`-like methods. This algorithm

TABLE I
Approaches to Calling Operations on JVM

|  | Reflection lookup | Direct dynamic calls | Method calls by the table | Indirect dynamic calls |
|---|---|---|---|---|
| **Boxing problem** | Only return value is boxed | Both arguments and return value are boxed | Only return value is boxed, or everything is boxed | Both arguments and return value are boxed, but some optimizations will be possible |
| **Fails if operation name doesn't match the method name** | Yes | No | No | No |
| **An extra parameter should be passed to the compiler** | No | No | Yes | No |
| **Performs tableswitch operations lookup** | Only if method can't be found | At each call | No | Only at compilation |

required a larger boxing allocation overhead; however, was much more stable and universal.

We now compare the bytecode generated by both generation algorithms in their decompiled forms: the legacy one in Fig. 1, and after our implementation in Fig. 2.

Both classes are generated from the expression `x+2` inside a `RealField` context, which implements `ExtendedField<Double>`, so both classes implement `Expression<Double>`. The access flags, class name pattern, declared methods, and type signatures (except the key of arguments map is changed to `Symbol` from `String`) are the same. The first difference is stored fields. The legacy bytecode generator emitted either one or two fields: the first which stores a reference to `Algebra` object, and the second which stores constants (as `Object[]`) required by the expression which cannot be placed into the class file's constant pool directly. In contrast, the new generator emits only one field—`constants`, which stores dynamic constants as well as Kotlin function objects produced by the `unaryOperationFunction` and `binaryOperationFunction` methods. The second difference is expression constructor: where originally a reference to `Algebra` serves as the receiver of all operations, the new one has elements of the constants array only. Both generated classes are constructed with reflection.

## VI. JavaScript Source Code Generation

Applying *Kotlin Multiplatform* to the created library, it was easy to port MST features to *Kotlin/JS* (Kotlin for JavaScript) and *Kotlin/Native* (Kotlin for Native) while supporting functionality similar to JVM dynamic compilation.

The development of Kotlin/JS was straightforward. The idea of storing functions instead of `Algebra` references was derived from the Java bytecode backend in which the generated function is assigned a similar constants array storing both constant values of the expression and function references to operations used in the expression.

As for tooling, *ESTree* [9] as JavaScript AST classes package and astring [10] as code generation framework were selected. The only implementation choice was between creating sources by appending fragments to a string or building an AST then rendering it.

The MST to JS compiler generates a function, then wraps it as a KMath `Expression`. There is an example of such a function in Listing 3.

```
var executable = function (constants, arguments) {
  return constants[1](constants[0](arguments, "x"), 2);
};
```

Listing 3. Example of generated JavaScript function.

## VII. WebAssembly IR Generation

*WebAssembly* [11] (also known as WASM) is an open standard defining a portable IR for executable programs, and in the context of this study, WebAssembly code generation was attempted; however, the WASM compiler path is much more limited than JVM bytecode or JS generation.

To facilitate the dynamic compilation, Kotlin/JS was used to prototype our backend implementation. In using it, several trade-offs and problems were encountered.

1) Due to slowness and lack of interoperability between JS and WASM, Kotlin/JS builtin mathematical functions (which are simply delegated to the JavaScript `Math` object) were unavailable as well as an opportunity to invoke KMath context functions.
2) Only `f64` and `i32` WASM types were supported, as `i64` is unavailable without an experimental V8 feature which maps `i64` to JavaScript's `bigint` type. `f32` was not available for a similar reason—JavaScript does not provide a type for single-precision floating-point values.
3) Basic mathematical functions for `f64` (e.g. `sin` and `cos`) required to support `RealField` operations were taken from `libm` (also known as `math.h` [12]), which was compiled to WASM and appended partially to the initial state of the WASM module. All other `f64` arithmetic is available with WASM opcodes.

```
(func $executable (param $0 f64) (result f64)
  (f64.add
    (local.get $0)
    (f64.const 2)
  )
)
```

Fig. 1. Example of emitted WASM IR in the WAT form.

```java
import java.util.*;

import scientifik.kmath.asm.internal.*;
import scientifik.kmath.expressions.*;
import scientifik.kmath.operations.*;

public final class AsmCompiledExpression_1073786867_0
    implements Expression<Double> {
  private final RealField algebra;

  public AsmCompiledExpression_1073786867_0(RealField algebra) {
    this.algebra = algebra;
  }

  public final Double invoke(Map<String, ? extends Double> arguments) {
    return (Double) algebra
        .add(((Double) MapIntrinsics
            .getOrFail(arguments, "x"))
            .doubleValue(), 2.0D);
  }
}
```

Listing 1. Legacy bytecode generation result (decompiled).

```java
import java.util.*;

import kotlin.jvm.functions.*;
import kscience.kmath.asm.internal.*;
import kscience.kmath.expressions.*;

public final class AsmCompiledExpression_45045_0
    implements Expression<Double> {
  private final Object[] constants;

  public AsmCompiledExpression_45045_0(Object[] constants) {
    this.constants = constants;
  }

  public final Double invoke(Map<Symbol, ? extends Double> arguments) {
    return (Double) ((Function2) constants[0])
        .invoke((Double) MapIntrinsics
            .getOrFail(arguments, "x"), 2);
  }
}
```

Listing 2. Current bytecode generation result (decompiled).

The backend described uses `binaryen` [13] library to simplify IR generation and perform various optimizations. While the upcoming *Kotlin/WASM* (Kotlin for WebAssembly) toolchain would have been a much more suitable target due to lower interoperability overhead, the project is still in early development as of this time.

## VIII. LLVM IR Generation

The *LLVM* [14] (Low-level Virtual Machine) compiler infrastructure project is a set of compiler and toolchain technologies designed around an IR which serves as a portable, high-level assembly language. LLVM was used as the backend of Kotlin/Native and was investigated as a possible expression compilation target.

However, the decision was made to forgo this feature for two reasons:

1) This generation target could not be made universal due to the difficulty of interoperation with Kotlin /Native as the host platform.
2) LLVM's monolithic scope and poor compilation performance, particularly with higher optimization levels, is unsuitable for the dynamic compilation, at least for primitive typed computations. All the previously mentioned IRs have a lightweight or built-in infrastructure for runtime code generation.

## IX. Benchmark Results

The new expression APIs were microbenchmarked. Two measurements are included in this paper.

**Environment data:**
- CPU: Intel Core i5 6400, 3.196 GHz, Skylake
- RAM: 15.977 GiB
- OS: Ubuntu 20.10 Groovy

All tested expressions API implementations calculate the following formula one million times using double-precision floating-point arithmetic:

$$2\,x + \frac{2}{x} - \frac{16}{\sin(x)}. \qquad (1)$$

**Java runtime:**
1) OpenJDK Hotspot (build 11.0.10+9-LTS)
2) OpenJDK GraalVM CE 21.0.0 (build 11.0.10+8-jvmci-21.0-b06)

The *JMH* [15] (Java Microbenchmark Harness) shipped within the kotlinx-benchmark [16] toolkit was used in throughput mode with 5 warm-ups and 5 plain iterations. JVM hosted measurements are presented in Table II.

TABLE II
JVM Hosted Measurements

|  | Description | Average throughput | |
|---|---|---|---|
|  |  | **Hotspot** | **GraalVM** |
| **functional** | Functional expression | 2.6 Hz | **4.003** Hz |
| **mst** | Interpretation of MST | **0.188** Hz | 0.177 Hz |
| **asm** | ASM compiled expression | 2.994 Hz | **4.196** Hz |
| **raw** | Statically compiled implementation in Kotlin | 4.012 Hz | **7.864** Hz |

**JS runtime:**
1) NodeJS 12.16.1 (V8 7.8.279.23-node.31)

JS hosted measurements are presented in Table III.

TABLE III
JS Hosted Measurements

|  | Description | Single shot time |
|---|---|---|
|  |  | **NodeJS** |
| **functional** | Functional Expression | 3.61 s |
| **mst** | Interpretation of MST | 254 s |
| **wasm** | WASM compiled expression | 4.22 s |
| **estree** | ESTree compiled expression | **3.55** s |
| **raw** | Statically written implementation in Kotlin | 4.78 s |

The benchmark data suggest the following implications:
1) GraalVM is generally faster than Hotspot.
2) Functional expressions are nicely optimized both by JVM and V8.

3) Raw expressions on JVM do almost no boxing conversions, and that's the reason they are the fastest.
4) WASM interoperability (passing of even primitive data) is very expensive.
5) Statically created expressions are significantly faster than dynamically compiled ones.
6) The JVM hosted benchmarks will need to be run on future JVM implementations after the *Project Valhalla* [17] release. Valhalla implements *JEP 218* [18], so boxing caused by generic interfaces can be eliminated by JVM.

## X. Conclusion

The research on the dynamical interpretation and code generation for generic algebras in KMath is a work in progress. Much work remains to be done to stabilize the API and improve performance. Still, current results show the potential of dynamic expression building even for performance-critical parts. While the ASM code generation did not provide a significant performance boost, it is still useful for research.

While MST representation was a side-product of this research, it proved to be a valuable tool on its own. As a syntactic tree that can be extended with support for future symbols, it is possible to use for simple symbolic computations. For example, there is experimental support for automatic differentiation based on Kotlin∇ [5].

## XI. Acknowledgment

The author would like to thank the members of the KMath development team (Alexander Nozik, Peter Klimai, and Roland Grinis) and Breandan Considine for the discussion and revision of the work.

The KMath project is developed in cooperation between MIPT and JetBrains Research.